\documentclass[preprint,authoryear,12pt]{elsarticle}

\newcommand{\Da}        {{\,\rm Da}}
\newcommand{\Pe}        {{\,\rm Pe}}
\newcommand{\Sh}        {{\,\rm Sh}}
\newcommand{\BB}[1]		{\mbox{\boldmath${#1}$}}
\newcommand{\Vq}		{{\BB{q}}}
\newcommand{\VD}		{{\BB{D}}}
\newcommand{\Vx}		{{\BB{x}}}
\newcommand{\nm}		{\mbox{nm}}

\newcommand{\mm}		{\mbox{mm}}
\newcommand{\cm}		{\mbox{cm}}
\newcommand{\m}			{\mbox{m}}

\newcommand{\M}			{\mbox{M}}
\newcommand{\s}			{\mbox{s}}
\newcommand{\FigWidth}	{240pt}
\newcommand{\FigWidthD}	{400pt}
\usepackage{graphicx,hyperref,movie15}
\usepackage{float}
 
\newfloat{movie}{thp}{lop}
\floatname{movie}{Movie}

\usepackage{amssymb,amsmath}

\journal{Earth and Planetary Science Letters}

\begin{document}

\begin{frontmatter}

\title{The initial stages of cave formation: Beyond the one-dimensional paradigm}


\author[uw]{Piotr Szymczak \corref{cor1}}
\ead{Piotr.Szymczak@fuw.edu.pl}

\author[uf]{ Anthony J.C. Ladd}
\ead{tladd@che.ufl.edu}

\cortext[cor1]{Corresponding author}

\address[uw]{Institute of Theoretical Physics, Warsaw University, Ho\.{z}a 69, 00-618, Warsaw, Poland}
\address[uf]{Department of Chemical Engineering, University of Florida, Gainesville, Florida 32611-6005, USA}

\address{}

\begin{abstract}
The solutional origin of limestone caves was recognized over a century ago, but the short penetration length of an undersaturated solution made it seem impossible for long conduits to develop. This is contradicted by field observations, where extended conduits, sometimes several kilometers long, are found in karst environments. However, a sharp drop in the dissolution rate of ${\rm CaCO_3}$ near saturation provides a mechanism for much deeper penetration of reactant. The notion of a ``kinetic trigger'' - a sudden change in rate constant over a narrow concentration range - has become a widely accepted paradigm in speleogenesis modeling. However, it is based on one-dimensional models for the fluid and solute transport inside the fracture, assuming that the dissolution front is planar in the direction perpendicular to the flow. Here we show that this assumption is incorrect; a planar dissolution front in an entirely uniform fracture is unstable to infinitesimal perturbations and inevitably breaks up into highly localized regions of dissolution. This provides an alternative mechanism for cave formation, even in the absence of a kinetic trigger. Our results suggest that there is an inherent wavelength to the erosion pattern in dissolving fractures, which depends on the reaction rate and flow rate, but is independent of the initial roughness. In contrast to one-dimensional models, two-dimensional simulations indicate that there is only a weak dependence of the breakthrough time on kinetic order; localization of the flow tends to keep the undersaturation in the dissolution front above the threshold for non-linear kinetics.
\end{abstract}

\end{frontmatter}


\section{Introduction}
\label{intro}

The notion that caves result from dissolution by an aqueous solution of carbon dioxide was already present in the works of Lyell and Thirria in the 1830s~\citep{Shaw2000}, but it was not until 100 years later that these ideas were developed mathematically. A quantitative model of the dissolution of a single limestone fracture was developed by~\citet{Weyl1958}, taking into account chemical kinetics and solute transport. His theory led to the paradoxical conclusion that water flowing through a limestone fracture becomes saturated with calcium ions over length scales of the order of centimeters, so that limestone caves should not exist at all~\citep{White1962}. A possible resolution of this paradox was proposed by~\citet{White1977}, who noted that the existence of large cave systems may be explained by the sharp drop in the dissolution rate of $\rm {CaCO_3}$ near saturation; this is frequently referred to as the ``kinetic trigger'' mechanism in the speleogenesis literature. Dissolution of blocks of calcite under laboratory conditions confirms that there is a rapid drop in reaction rate near saturation, apparently because of impurities in natural calcite~\citep{Plummer1976,Dreybrodt1990,Palmer1991,Svensson1992}.

Several calculations~\citep{Dreybrodt1990,Palmer1991,Groves1994,Dreybrodt1996} have shown that the kinetic trigger hypothesis allows for the development of deep conduits, but they are all based on a one-dimensional model of fracture dissolution. The initial fissure is approximated by two parallel planes, and all the relevant variables (aperture, fluid volume flux, and solute concentration), depend only on the distance from the inlet (see Fig.~\ref{fig1}). However real fractures are never one-dimensional; in this paper we show that even a tiny variability in fracture aperture inevitably leads to an instability at the reaction front, which then breaks up into highly localized regions of dissolution. Thus, the premise that a smooth fracture will open uniformly across its width, which underlies current models of speleogenesis, is faulty. A correct description of fracture dissolution must include the variation in aperture across the lateral direction as well. 

\section{Breakthrough times in one and two dimensions}\label{model}

In this paper we reevaluate the standard mathematical model for the early stages of cave formation, describing the dissolution of a calcite fracture by surface water draining through it to a lower-lying water table. We analyze the coupled equations for fluid flow, reactant transport and surface dissolution to show that the evolution of the fracture aperture is an inherently two-dimensional process, even when the initial aperture field is spatially uniform.

\subsection{Flow and transport: 1D model}\label{1d}

\begin{figure}
\center \includegraphics[width=\FigWidth]{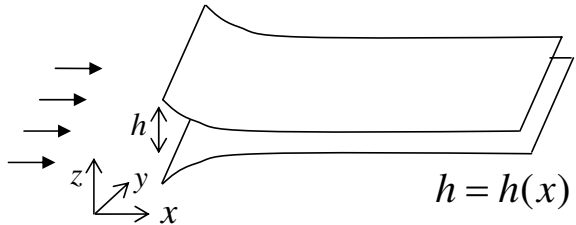}
\caption{Dissolution of a one-dimensional fracture; fluid flow is in the $x$ direction and the fracture surfaces dissolve in the normal ($z$) direction. The fracture surfaces can be assumed to be located at $\pm h/2$.}\label{fig1}
\end{figure} 

In studies of fracture dissolution, and particularly in theoretical 
investigations of cave formation, a one-dimensional model of a single fracture is frequently used \citep[e.g.][]{Dreybrodt1990,Groves1994,Dreybrodt1996,Dijk1998}. 
 In this model, schematically represented in Fig.~\ref{fig1}, the fracture aperture $h(x,t)$ is assumed to depend on a single spatial variable, the distance from the inlet. The flow rate, $q(t)$, is then independent of position,
\begin{equation}\label{eq:1dflow}
 q(t) = \frac{{\Delta p}}{r(t)}, \ \ \ \ r(t) = 12\mu \int\limits_0^L {\frac{{dx}}{{{h(x,t)^3}}}},
\end{equation} 
where $\Delta p$ is the difference between the inlet and outlet pressures, $L$ is the fracture length along the flow ($x$) direction, and $r$ is the integrated flow resistivity. 

The concentration of $\rm Ca^{2+}$ ions in the fracture, $c(x,t)$, is described by a convection-diffusion equation
\begin{equation}\label{eq:1dtransport}
 q \frac{dc}{dx} - \frac{d}{dx}\left(h D_{xx}\frac{dc}{dx}\right) = 2 R(c),
\end{equation} 
where $D_{xx}$ is the dispersion coefficient and $R(c)$ is the reactive flux from the dissolving calcite. The factor of two in the dissolution rate comes from combining erosion at the upper and lower fracture surfaces. It is assumed that the inlet stream is pure water, $c(0,t) = 0$, and the outlet stream is a saturated solution, $c(L,t) = c_{sat}$. Finally, the aperture evolution is determined from the local reactive flux
\begin{equation}\label{eq:1dgeom}
c_{sol} \frac{dh}{dt} = 2 R(c)
\end{equation} 
where $c_{sol}$ is the molar concentration of the solid phase. 

\subsection{Dissolution and breakthrough: 1D model}\label{DB1d}

\begin{figure}
\center \includegraphics[width=\FigWidth]{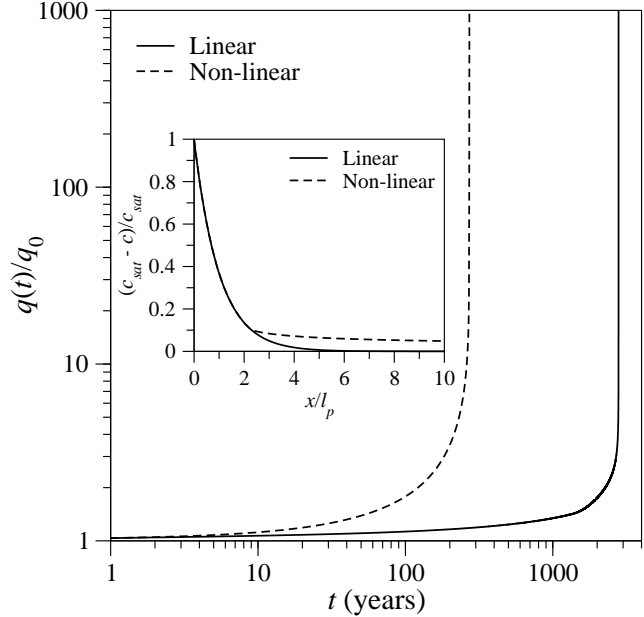}
\caption{One-dimensional dissolution of a $20{\m}$ fracture with a uniform initial aperture $h(x) = h_0 = 0.2{\mm}$; $\Pe = 100$ and $\Da = 5\times10^{-4}$. The volumetric flux $q(t)$, scaled by the initial flux in the fracture $q_0$, is shown for linear and non-linear kinetics. The inset shows the concentration profile from Eqs.~\eqref{eq:1dtransport} and~\eqref{eq:R}, with $c_{th} = 0.9c_{sat}$.}\label{fig2}
\end{figure}

Early theories of karstification assumed linear dissolution kinetics~\citep{Weyl1958,White1962},
\begin{equation}
\label{eq:R1}
R(c) = k(c_{sat}-c).
\end{equation}
In this case the undersaturation decays exponentially into the fracture, $c_{sat}-c \sim e^{-x/l_p}$, with a penetration length,
\begin{equation}\label{eq:lp}
 l_p = q /2k.
\end{equation}
Dissolution is restricted to a narrow region near the inlet, $x \sim l_p$, which limits the growth of conduits. In fractured carbonate formations $l_p$ is typically less than a meter \citep{White1962,Atkinson1968,Dreybrodt1990}. \citet{White1977} proposed that field observations of extended conduits in karst environments, sometimes several kilometers long, might best be explained by a sharp drop in mineral dissolution rate as saturation is approached. This ``kinetic trigger'' mechanism can be modeled as a switch from linear to higher order kinetics at a threshold concentration $c_{th}$:
\begin{equation}
\label{eq:R}
R(c) = \left\{ \begin{array}{cc}
 \!\!\!\!k(c_{sat}-c), & c<c_{th} \\
 k_n(c_{sat}-c)^n, &c>c_{th}. \\ 
 \end{array} \right.
\end{equation}
A non-linear kinetic law ($n > 1$ in Eq.~\eqref{eq:R}) results in an algebraic decay of the concentration profile at large distances from the inlet, $c_{sat}-c \sim x^{1/1-n}$, and the fracture opens over its entire length.  Concentration profiles for linear and non-linear kinetics are illustrated in the inset to Fig.~\ref{fig2}. In these (and subsequent) calculations we take typical parameter values for calcite dissolution: $n=4$, $k = 2.5 \times 10^{-5} \cm \s^{-1}$  and $c_{sat} = 2 \times 10^{-6} \M \cm^{-3}$~\citep{Dreybrodt1996}. The reaction rate $k_4 = k (c_{sat} - c_{th})^{-3}$ is adjusted so that $R(c)$ is continuous at $c = c_{th}$, and the threshold concentration itself was set to $0.9 c_{sat}$. The molar concentration in the solid phase was taken as $0.027 \M \cm^{-3}$, based on the mass density of pure calcite.

In this paper we take the initial fracture aperture $h_0 = 0.2 \mm$, which means that the product of P\'eclet number, $\Pe = q_0/D$, and Damk\"ohler number, $\Da = k h_0/q_0$, is about $0.05$. Here $q_0$ is the initial value of the volumetric flux in the fracture and the solute diffusion constant $D \approx 10^{-5} \cm^2\s^{-1}$.

Figure~\ref{fig2} shows typical breakthrough curves for linear and non-linear kinetics, obtained by solving Eqs.~\ref{eq:1dflow}--\ref{eq:1dgeom} numerically (see~\ref{sim} for details). The volumetric flux in the initial fracture is about $10^{-3} \cm^2\s^{-1}$, corresponding to an hydraulic gradient of $0.01$. The penetration length, $l_p \approx 20 \cm$, is much less than the length of the fracture ($L = 20 \m$). Thus with linear kinetics the opening of the fracture aperture is slow, with the breakthrough time scaling exponentially with the length of the fracture. Breakthrough occurs much earlier in the non-linear case, because the slower kinetics allow a deeper penetration of reactant into the fracture~\citep{White1977,Dreybrodt1990}. The  time scale for dissolution can be characterized by the growth of the fracture aperture at the inlet, $h(0,t) = h_0 + 2k t c_{sat}/c_{sol}$, which is independent of the kinetic model and the dimensionality of the fracture. We define
\begin{equation}
\label{eq:tau}
\tau = h_0 c_{sol} /2 k c_{sat}
\end{equation}
as the time it takes for the initial fracture aperture at the inlet to double. For an $0.2 \mm$ fracture, $\tau = 5.4 \times 10^6 \s$ or approximately $0.17$ years.

\subsection{Flow and transport: 2D model}\label{2d}

Although one-dimensional models are simple and mathematically tractable, laboratory experiments~\citep{Durham2001,Gouze2003,Detwiler2003} have shown that in most cases fracture dissolution is non-uniform in the direction transverse to the flow, with highly localized two-dimensional dissolution patterns~\citep{Hanna1998}. The non-linear dynamics underlying this more complicated behavior can be probed using {\it depth-averaged} models~\citep{Cheung2002,Detwiler2007}, in which the fluid velocity and reactant concentration are still averaged over the fracture aperture ($z$ direction in Fig.~\ref{fig1}), but can vary in the lateral ($y$) direction. Coupled equations for the fluid volume flux $\Vq(x,y,t)$, depth-averaged concentration of dissolved solids $c(x,y,t)$, and aperture $h(x,y,t)$ are solved simultaneously:
\begin{equation}\label{eq:2D}
\begin{array}{cc}
\nabla \cdot \Vq = 0, ~~ \Vq =  -\dfrac{h^3}{12}\dfrac {\nabla p}{\mu}, \\
\Vq \nabla \cdot c - \nabla \cdot  (h \VD \cdot \nabla c) = 2 R(c) \\
c_{sol} \partial_t h  =  2 R(c),
\end{array}
\end{equation}
where $\VD(h)$ is the solute dispersion tensor. Details of our numerical solution of Eq.~\eqref{eq:2D} can be found in~\ref{sim}.

In Sec.~\ref{instab} we will show mathematically that the one-dimensional dissolution profiles described by Eqs.~\eqref{eq:1dflow}--\eqref{eq:1dgeom}, are unstable. Uniform concentration profiles inevitably break down into more complex dissolution patterns, which can be described by Eq.~\eqref{eq:2D}. Simulations based on Eq.~\eqref{eq:2D} have proven to be successful in predicting the large scale dissolution patterns in artificial fractures~\citep{Detwiler2007}, although at small scales, comparable to the  correlation length in the aperture field, the dissolution is more uniform than the experimental observations. For such fine details three-dimensional simulations are necessary, including the variation in flow and concentration fields across the aperture~\citep{Szymczak2009}.

\subsection{Dissolution and breakthrough: 2D model}\label{DB2d}

\begin{figure}
\center \includegraphics[width=\FigWidth]{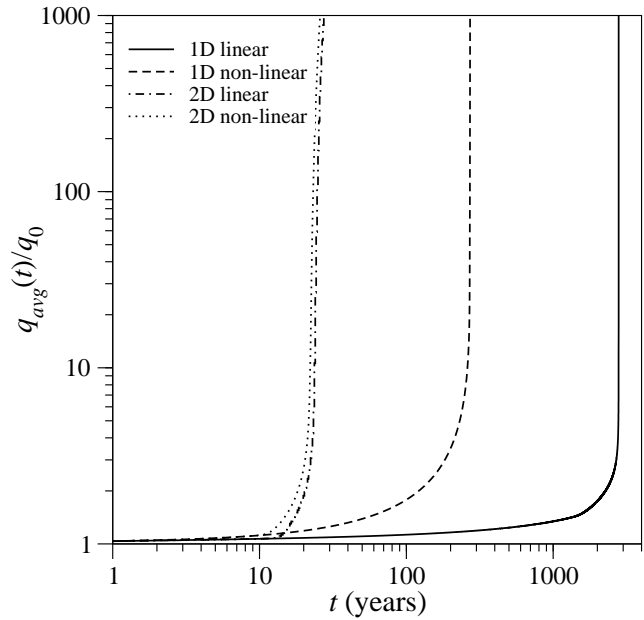}
\caption{Dissolution of a smooth fracture $20\m \times 10\m \times 0.2{\mm}$; the parameters $\Pe = 100$ and $\Da = 5\times10^{-4}$. One-dimensional simulations of the volumetric flux $q(t)$ are compared with two-dimensional simulations of the average flux $q_{avg}(t) = W^{-1}\int_0^W q_x(x,y,t) dt$. The fluxes are scaled by the initial flux in the fracture $q_0$.}\label{fig3}
\end{figure}

We repeated the one-dimensional simulations shown in Fig.~\ref{fig2} using a two-dimensional fracture, $20 {\m} \times 10 {\m}$, superimposing a small random aperture field, with a variance $\sigma = 20 {\nm}$, over the original uniform aperture ($0.2 {\mm}$). Time-dependent flow rates from 1D and 2D simulations are compared in Fig.~\ref{fig3}. Two-dimensional simulations show a significant reduction in the breakthrough time in comparison with 1D models, in agreement with earlier results~\citep{Hanna1998,Szymczak2009}. Interestingly, the added spatial dimension has a larger effect on the breakthrough time than the kinetic trigger. In the one-dimensional simulations, non-linear kinetics reduces the breakthrough time by an order of magnitude, from 2800 years to 270 years. However, in the two-dimensional simulations, the breakthrough time is less than 30 years, regardless of the kinetics. In fact the non-linear kinetics makes only a small difference here, reducing the breakthrough time from about 28 years (linear) to 26 years (non linear).

\begin{figure}
\center\includegraphics[width=\FigWidthD]{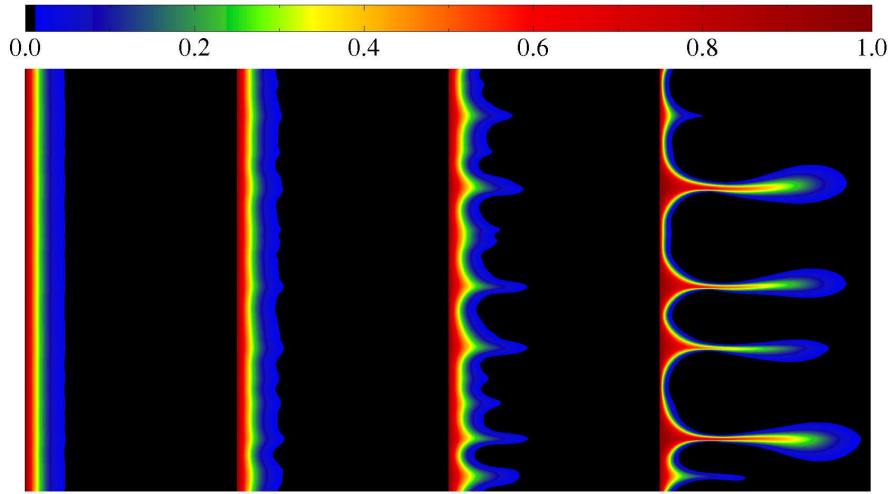}
\caption{Concentration profiles in a dissolving fracture. Contour plots of the normalized undersaturation, $(c_{sat}-c)/c_{sat}$, are plotted at successive times (from left to right); $t = 25 \tau$, $t = 50 \tau$, $t = 75 \tau$, and $t = 100 \tau$. The flow direction is from left to right; the P\'eclet and Damk\"ohler numbers are $\Pe = 100$ and $\Da = 5\times10^{-4}$. A video of the complete time sequence is included as an ancillary file to this article and also accessible at \url{http://www.fuw.edu.pl/~piotrek/c.mpg}.}~\label{fig4}
\end{figure}

The reason why dimensionality is crucial can be seen in the concentration maps shown in Fig.~\ref{fig4}. The initially planar dissolution front breaks down into increasingly focused regions of dissolution, which rapidly advance into the fracture, causing breakthrough at much earlier times than in the homogeneous (one-dimensional) case~\citep{Hanna1998,Szymczak2009}. Flow focusing keeps the concentration in the advancing front high, above the threshold for non-linear kinetics; this explains the relatively small difference in breakthrough times in the two-dimensional simulations.

In the course of extensive numerical investigations of fracture dissolution, we noticed that a planar dissolution front was always unstable if the fracture length, $L$, was sufficiently large in comparison with the penetration length, $l_p$. Our simulations suggested that on geophysical scales, where $L \gg l_p$ dissolution will be inherently unstable. Mathematical analysis, described in~\ref{Lsa}, confirms that fracture dissolution is always two-dimensional, and cannot be usefully approximated by one-dimensional models.

\section{Instability of dissolution in a uniform aperture fracture}\label{instab}

The stability of a uniform dissolution front can be investigated from a minimal model, assuming linear kinetics, and neglecting the dispersive term in Eq.~\eqref{eq:2D}:
\begin{equation}\label{eq:2Dmin}
\begin{array}{cc}
\nabla \cdot \Vq = 0, ~~ \Vq =  -\dfrac{h^3}{12}\dfrac {\nabla p}{\mu}, \\
\Vq \cdot \nabla c = 2k(c_{sat}-c) \\
c_{sol} \partial_t h  =  2k(c_{sat}-c).
\end{array}
\end{equation}
In Sec.~\ref{DB2d} we showed (Fig.~\ref{fig3}) that once a two-dimensional model of fracture dissolution is adopted, variations in kinetic order are of only quantitative rather than qualitative importance to the predicted breakthrough time. Moreover, on the scale of penetration length, the ratio of the diffusive flux, $Dhc/l_p^2$, to the convective flux, $qc/l_p$, is of the order of $\Pe^{-1} \Da$. This parameter is small in typical limestone fractures, as shown by a consideration of fracture apertures and hydraulic gradients (\ref{DaPe}); thus Eq.~\eqref{eq:2Dmin} contains all the essential physics of the instability. However, in gypsum karst the reaction rate is much higher ($\Da > 1$) and the penetration length equivalently smaller; in this case diffusive flux is significant and should be included, as in Eq.~\eqref{eq:2DlinD}.

In the initial stages of dissolution, when the penetration length is much less than the length of the fracture, $l_p \ll L$, the one-dimensional flow and concentration fields are essentially time independent,
\begin{equation}\label{eq:1D}
q(t) \approx q_0 = -\frac{h_0^3 p_0^\prime}{12 \mu}, ~~ c(x,t) \approx c_0(x) = c_{sat}(1- e^{-x/l_p}).
\end{equation}
The pressure gradient $p_0^\prime = \partial_x p$ is constant far from the inlet and $q$ is independent of $x$ by continuity; the subscript $0$ is used to denote initial values. The aperture grows linearly in time, with a rate that decays exponentially with the distance from the inlet,
\begin{equation}\label{eq:h1D}
h(x,t) = h_0 \left(1 + \frac{te^{-x/l_p}}{\tau}\right).
\end{equation}
The flow rate and concentration field remain constant in time until the penetration is deep enough to perturb the far field pressure gradient and increase the flow rate in the fracture.

\begin{figure}
\center \includegraphics[width=\FigWidth]{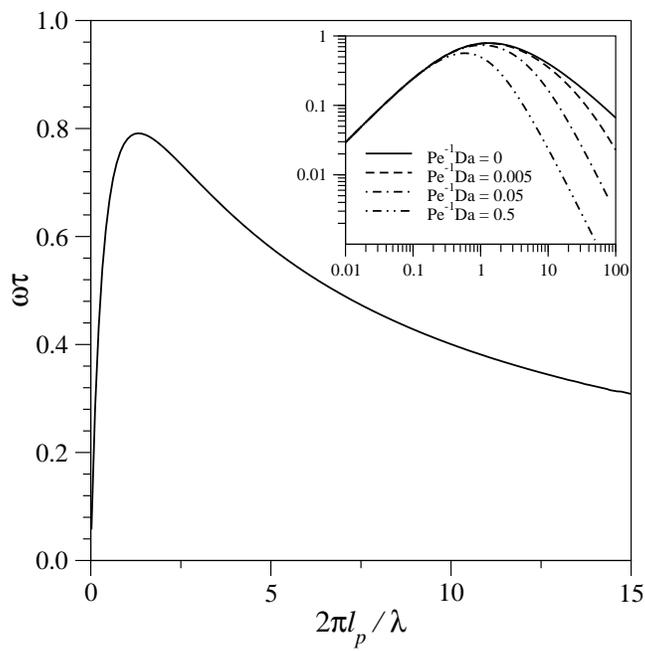}
\caption{Growth rate of the instability in a smooth fracture. The time scale $\tau = h_0/2k\gamma$. The inset figure shows the growth rate over a larger range of wavelengths, including the effects of lateral diffusion as measured by the product $\Pe^{-1}\Da$.}\label{fig5}
\end{figure}

However, an initially smooth fracture does not evolve according to Eq.~\eqref{eq:h1D}; in fact the one-dimensional solution is exponentially unstable to perturbations in $h$, $\delta h(x,y,t) \sim e^{\omega t}$, which rapidly outrun the uniform front suggested by Eq.~\eqref{eq:h1D}. A linear stability analysis of Eqs.~(\ref{eq:2Dmin}) leads to a dispersion relation for the growth rate $\omega$ as a function of the wavelength of the perturbation $\lambda$, as shown in Fig.~\ref{fig5}. The particulars of the analysis can be found in~\ref{Lsa}. Here we just mention one subtle but important detail; the base state, Eq.~\eqref{eq:h1D}, is itself time-dependent. The stability of nonautonomous systems is in general a very difficult problem~\citep{Farrell1996}; here however we follow a relatively simple, approximate approach~\citep{Tan1986} in which the base state is frozen at a particular time and the growth rate is determined as if the base state were time-independent (the quasisteady-state approximation). The dispersion curve in Fig.~\ref{fig5} was obtained by freezing the base state at $t_0=0$. We will examine this approximation in more detail in Sec.~\ref{numres}.

The most striking result of the linear stability analysis is that there is a maximal growth rate, $\omega_{max} = 0.79 \tau^{-1}$, at a wavelength $\lambda_{max} = 4.74 l_p$ (Fig.~\ref{fig5}).  An individual fracture can therefore develop a strongly heterogeneous permeability during dissolution, with an inherent length scale that depends on the kinetics and flow rate (via $l_p$), but not the initial topography. Furthermore, there is no lower limit to the reaction rate for unstable dissolution. Only the wavelength and penetration depth are affected, scaling with the inverse of the Damk\"ohler number.

In laboratory experiments it is frequently the case that the sample length is less than $l_p$, in which case uniform dissolution is observed~\citep{Fredd1998}. But on scales of geophysical relevance, dissolution of carbonate rocks will always be unstable; calcite for example would have characteristic wavelengths $\lambda_{max} \sim 1 \m$, while in dolomite $\lambda_{max} \sim 10\m$. It is worth noting that the instability in the dissolution front continues to grow in systems confined to widths smaller than $l_p$, although at a reduced rate. Figure~\ref{fig5} shows that there is a long tail to the dispersion relation at short wavelengths, with a weak power-law dependence $\omega ~\sim \lambda$. Thus in narrow fractures, where the width $W < \lambda_{max}$, a single channel develops which contains all the flow in the fracture. 

The instability analysis in~\ref{Lsa} can be generalized to include lateral diffusion (in the $y$ direction), which is expected to play a more important role in the dynamics than axial diffusion because axial transport is usually convection dominated. On scales comparable to the penetration length, the relative magnitude of diffusive and convective fluxes is given by the parameter $\Pe^{-1}\Da$, as can be seen in Eq.~\eqref{eq:2DlinD}. The inset to Fig.~\ref{fig5} shows the dispersion relation for several different values of the product $\Pe^{-1}\Da$. Lateral diffusion reduces the growth rate of short wavelength perturbations and shifts $\lambda_{max}$ towards
longer wavelengths. However, it does not prevent the instability developing and the growth rate remains positive for all wavelengths.

\section{Numerical simulations of the instability}\label{numres}

\begin{figure}
\center \includegraphics[width=\FigWidthD]{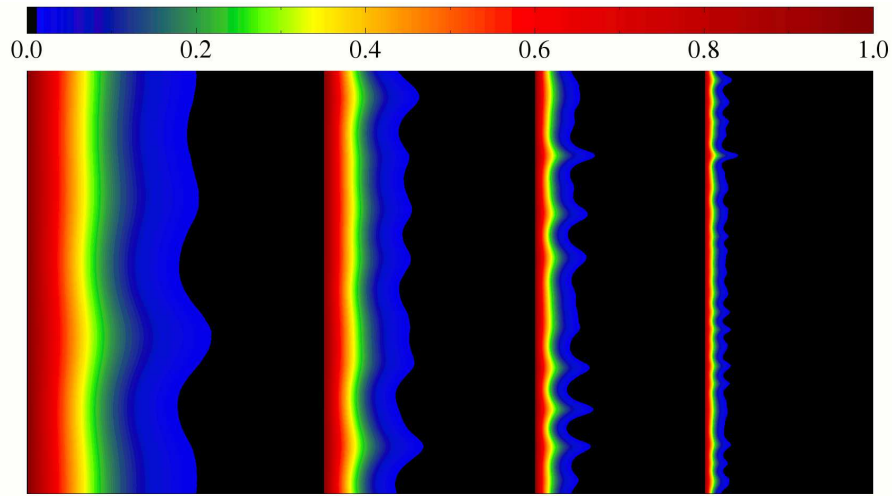}
\caption{ Concentration profiles of a dissolving fracture. Contour plots of the normalized undersaturation, $(c_{sat}-c)/c_{sat}$, are plotted for different Damk\"ohler numbers (from left to right); $\Da = 1.25 \times 10^{-4}$, $2.5 \times 10^{-4}$, $5.0 \times 10^{-4}$, and $\Da = 10^{-3}$. The product of P\'eclet and Damk\"ohler numbers is $0.05$ in each case. The flow is from left to right and the time $t = 50 \tau$. The variance in the aperture was reduced to $10^{-5} h_0$ so that the sinusoidal perturbations in the front are more easily visible at larger $\Da$.}~\label{fig6}
\end{figure}

\begin{figure}
\center \includegraphics[width=\FigWidth]{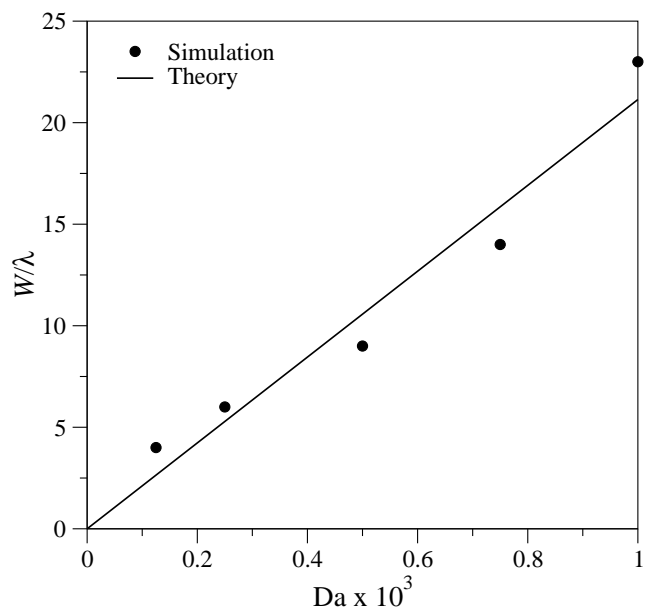}
\caption{Wavelength dependence of the dominant mode as a function of Damk\"ohler number, $\Da$. The solid circles show the number of wavelengths in a system of fixed width $W = 10\m$, determined by a Fourier analysis of the concentration profile. The solid line is the theoretical prediction based on wavelength of the fastest-growing mode, $\lambda_{max}=4.74 l_p$}~\label{fig7}
\end{figure}

We have confirmed the key predictions of the instability analysis
by numerically solving Eqs.~\eqref{eq:2D} with linear reaction kinetics. In comparison with the minimal model considered in the previous section, we have here included solute dispersion and the effect of diffusion on the dissolution kinetics, Eq.~\eqref{eq:k_eff}. Beginning from a uniform fracture ($h_0 = 0.2 \mm$), with a small random roughness superimposed ($\sigma = 10^{-4}h_0$), we see a single sinusoidal mode developing in the dissolution front, as illustrated in Fig.~\ref{fig4}. The wavelength and front thickness are inversely proportional to Damk\"ohler number, as shown in Fig.~\ref{fig6}, confirming that penetration length, $l_p = h_0/2 \Da$, is the only important length scale in the early stages of dissolution. Concentration profiles, such as those illustrated in Fig.~\ref{fig6}, can be Fourier transformed in the lateral ($y$) direction to determine the wavelength of the instability as a function of the Damk\"ohler number. An analysis of the number of wavelengths, $W/\lambda_{max}$, in a system of width $W = 10\m$ confirms that the fastest growing modes have a wavelength close to the theoretical prediction $\lambda_{max} = 4.74 l_p$ ($l_p = h_0/2\Da$), as can be seen in Fig.~\ref{fig7}.

\begin{figure}
\center \includegraphics[width=\FigWidth]{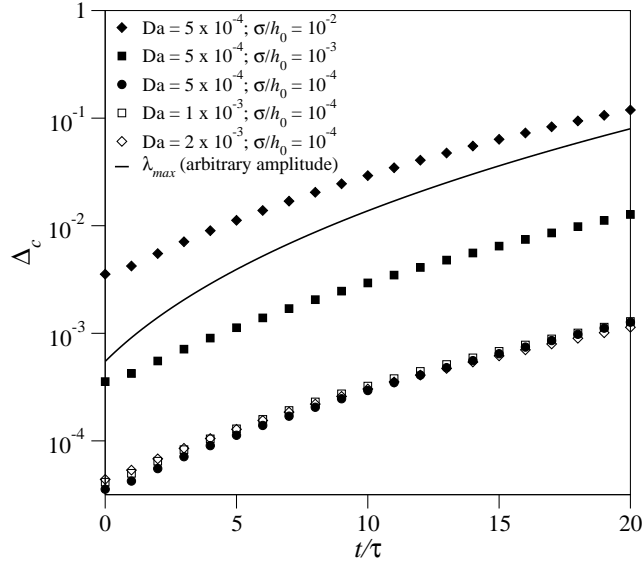}
\caption{Time-dependent fluctuations in concentration for different values of the Damk\"ohler number, $\Da$, and roughness, $\sigma/h_0$; $\sigma$ is the variance in the aperture. The solid line shows the growth of the mode with an initial wavelength of  $\lambda_{max}$, based on a numerical integration of the linearized equation, Eq.~\eqref{eq:flin}}~\label{fig8}
\end{figure}

The growth rates of the instability can be obtained by analyzing fluctuations in concentration field,
\begin{equation}\label{eq:cfluc}
\Delta_c (x,t) = \sqrt{\sum_y [c(x,y,t) - {\bar c}(x,t)]^2}.
\end{equation}
We take the maximum value of $\Delta_c (x,t)$ along the fracture, $\Delta_c(t)$, as a measure of the amplitude of the perturbation at time $t$. Results for $\Delta_c(t)$, presented in Fig.~\ref{fig8}, confirm that the instability is initiated as soon as the undersaturated solution enters the fracture, and that the growth is indeed approximately exponential in time. In reduced time units ($t/\tau$) the growth rate is independent of Damk\"ohler number and roughness; only the amplitude of the fluctuations is affected by the initial variability in the fracture aperture. For comparison, we also plot the solution of the linearized initial value problem, Eq.~\eqref{eq:flin}, which includes the time dependence of the base aperture field, Eq.~\eqref{eq:h1D}. We take the eigenfunction, Eq.~\eqref{eq:c}, corresponding to the fastest-growing eigenmode, $u = 2\pi l_p/\lambda_{max}$, as the initial condition. Initially, fluctuations in a random aperture field grow more slowly than the fastest-growing mode, but after a while ($\approx 10 \tau$) a single mode is dominant, and the growth rates become similar.

\begin{figure}
\center \includegraphics[width=\FigWidth]{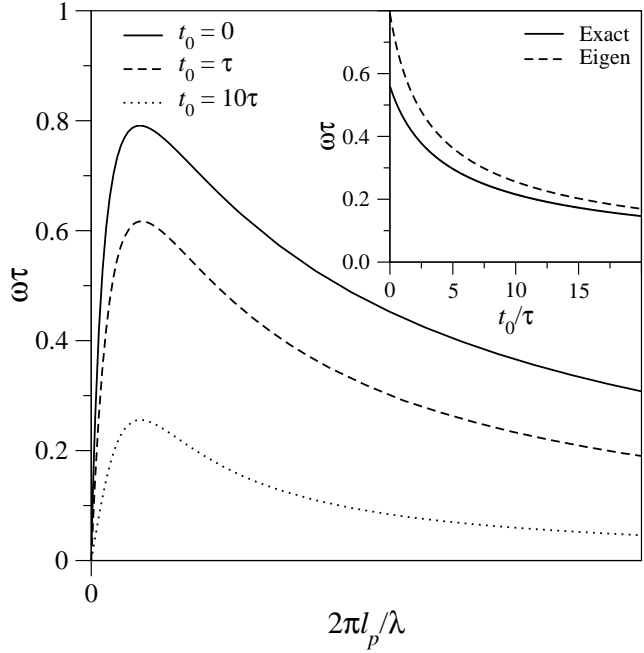}
\caption{Growth rate of the instability in a smooth fracture with the base state frozen  at different times: $t_0 = 0$ (solid line), $t_0=\tau$ (dashed line), and $t_0 = 10\tau$ (dotted line). The inset figure shows the growth rate of the mode with an initial wavelength of $\lambda_{max} = 4.74l_p$, for different times: the linearized solution (solid line), including the time-dependence of the base state, Eq~\eqref{eq:flin}, is compared with the eigensolution analogous to Eq.~\eqref{eq:disp} (dashed line)}\label{fig9}
\end{figure}

The linear stability analysis summarized in Fig.~\ref{fig5} predicts an exponential growth of the instability. On the other hand, the numerical results in Fig.~\ref{fig8} indicate that the growth is less than exponential. This is because the time dependence of the base state weakens the instability as the fracture inlet opens up. Figure~\ref{fig9} shows the dispersion relation obtained from the quasi-steady state approximation, Eq.~\eqref{eq:flin0}, for different base states, frozen at $t_0 = 0$ (as in Fig.~\ref{fig5}), $t_0 = \tau$, and $t_0 = 10\tau$; the growth rate is reduced across the whole spectrum of wavelengths as time goes on. However, the peak growth rate remains at almost the same wavelength, independent of time, so a single mode ($\lambda \approx \lambda_{max}$) predominates until the onset of non-linear growth.

The quasi-steady-state approximation, Eq.~\eqref{eq:flin0}, used to calculate the dispersion curves in Fig.~\ref{fig9} (and Fig.~\ref{fig5}) overestimates the growth rate when compared to the exact linearization in Eq.~\eqref{eq:flin}. The solid line in the inset to Fig.~\ref{fig9} shows the growth rate of the mode with an initial wavelength of $\lambda_{max}$ as a function of time; in essence the time derivative of the solid line in Fig.~\ref{fig8}. The dashed line is the growth rate for the same wavelength ($\lambda_{max}$) obtained by freezing the base state at successive moments of time. The discrepancy is about $40\%$ at $t_0 = 0$ and decreases thereafter, which agrees with the observation of~\citet{Tan1986} that the quasi-steady-state approximation becomes exact in the long-time limit.

\section{Discussion}

The dissolution of an initially smooth fracture is one dimensional at early times, but soon tiny random variations in fracture aperture lead to the development of small perturbations in the front. The strong wavelength selection (Fig.~\ref{fig5}) means that after a fairly short time, approximately $1-2$ years in our model calcite fracture, a single mode becomes dominant, with a wavelength that is about 5 times the penetration length (Fig.~\ref{fig7}). As the perturbation becomes stronger, distinct channels emerge from the front as shown in Fig.~\ref{fig4}. The  \href{http://www.fuw.edu.pl/~piotrek/c.mpg}{video} illustrates this process dynamically; the shorter channels are drained of reactant and cease to grow~\citep{Fredd1998}. Thus the spacing between long conduits is not determined solely by the initial wavelength of the instability but also by the non-linear dynamics of interacting channels. Eventually all the flow is localized within a few active channels. In this paper we have shown that this is the expected behavior on sufficiently large scales.

The notion of a kinetic trigger is an attractive explanation for the existence of long conduits in calcite formations, where there is an abundance of evidence for a transition from linear to non-linear kinetics~\citep{Plummer1976,Palmer1991,Svensson1992}. However, in gypsum there is good reason to think that the dissolution rate is roughly linear in the undersaturation~\citep{Colombani2007,Colombani2008}, although it has been suggested that there is a changeover to non-linear kinetics in natural gypsum as well~\citep{Jeschke2001}. One-dimensional models of gypsum dissolution do not develop long conduits for typical fracture apertures and flow rates~\citep{Raines1997}, unless a kinetic trigger with a high-order ($n=4$) rate law is again invoked~\citep{Dreybrodt2002,Romanov2003}.

The penetration length, Eq.~\eqref{eq:lp}, increases with time due to the increasing flow rate, which introduces a feedback mechanism that produces the characteristic upturn in $q(t)$ near breakthrough (Figs~\ref{fig2} and~\ref{fig3}). In one-dimensional models an increase in flow rate can only arise from the growth of the mean aperture. However in two-dimensions {\it flow focusing} provides a mechanism for a localized increase in flow rate; longer channels drain the flow from neighboring regions of the fracture, promoting their own growth at the expense of their neighbors~\citep{Fredd1998,Hanna1998}. The process can be understood by considering the pressure fields in channels of different lengths~\citep{Szymczak2006}. Given enough time, only a single channel remains and at that point the flow focusing mechanism stalls; the flow rate in the remaining channel can now only grow by extending its length. Thus there is a maximum flow rate that a single channel can acquire by flow focusing, $q_{max} \sim q_0 W/w$, where $w$ is the width of a single dissolution channel or wormhole. These observations suggest that the aspect ratio of the fracture and its orientation with respect to the flow may also play an important role in determining the breakthrough time. If $W \ge L$ the drainage basin is large and breakthrough will be rapid (as in the  \href{http://www.fuw.edu.pl/~piotrek/c.mpg}{video}), but if $W \ll L$ the flow focusing may stall with the dissolution front far from the outlet; in this case breakthrough will be dependent on reaction kinetics.

Discrete fracture networks are widely used in reservoir modeling~\citep{Long1982,Siemers1998} and in the assessment of dam safety~\citep{Dreybrodt2002,Romanov2003}, because they provide an effective means to describe localized regions of high permeability. However, each fracture within the network is modeled  at the one-dimensional level, and the possibility of spatially localized dissolution within an individual fracture in the network is not taken into account. In fractured carbonates an instability in the dissolution front can reduce the breakthrough time in an individual fracture by orders of magnitude, thereby completely changing the hydraulic properties of the global fracture network. It is therefore important to develop a model of evolving fracture permeability that includes the inherent heterogeneity of the dissolution process. Although the initial wavelength of the developing instability is controlled by the reaction kinetics and flow rate, interactions between the developing channels play the dominant role in the  long-time evolution of the fracture aperture field. Water throughout most of the fracture is saturated with calcium ions; only in the narrow regions formed by the active channels is there any significant undersaturation. Thus, an adequate understanding of how single channels advance (Fig.~\ref{fig4}  and  \url{http://www.fuw.edu.pl/~piotrek/c.mpg}{video}) would open up the possibility of new field-scale models of fracture dissolution, describing the growth of individual channels coupled together by long-range pressure fields. This could make it feasible to include heterogeneous dissolution of individual fractures in reservoir modeling, since the pressure fields around long thin channels can be rapidly calculated using techniques of conformal mapping~\citep{Gubiec2008}.

Our analysis differs from previous work on instabilities in porous media~\citep{Chadam1986,Sherwood1987,Hinch1990} in that we take a time-dependent aperture as the base state, rather than a propagating one-dimensional dissolution front. In a porous rock all the soluble material eventually dissolves and therefore a time-independent porosity profile can steadily advance into the material. However, in a fracture the aperture profile evolves along the whole length, with an effectively unlimited increase in aperture near the inlet.

\section{Conclusions}

The main conclusion from this work is that the dissolution of a single fracture is inherently two-dimensional; the one-dimensional solutions frequently used in models of cave formation~\citep{Dreybrodt1996} are unstable to infinitesimal perturbations. This means that the kinetic trigger mechanism is not a prerequisite for the development of long conduits. Although the development of instabilities in fracture dissolution has been demonstrated by numerical simulation~\citep{Hanna1998,Cheung2002,Szymczak2006,Szymczak2009}, this is the first time, to our knowledge, that the equations for fracture dissolution have been shown mathematically to have unstable solutions. We have further shown that the instability develops with a well-defined wavelength, which depends on reaction kinetics and flow rate but is insensitive to the initial roughness of the fracture. Subsequently, the localized regions of dissolution advance in the fracture in ways that may eventually be understood in terms of modified models of Laplacian growth.

\section*{Acknowledgments}
This work was supported by the Polish Ministry of Science and Higher Education (Grant No. N202023 32/0702), and by the US Department of Energy, Chemical Sciences, Geosciences and Biosciences Division, Office of Basic Energy Sciences (DE-FG02-98ER14853).

\appendix

\section{Simulation method}\label{sim}

The depth-averaged equations, Eq.~\eqref{eq:2D}, were solved on a uniform grid with a resolution of $1 \cm$; further grid refinement did not affect the results significantly. Periodic boundary conditions were imposed in the lateral ($y$) direction, while the inlet and outlet conditions were the same as in the one-dimensional case, Sec.~\ref{1d}. We included Taylor dispersion in the axial diffusion, $D_{xx} = D+q_x^2/210D$, but its effect on the erosion rate was negligible; in the lateral direction we took $D_{yy}=D$. One-dimensional calculations used the same code, but with a single grid point in the lateral ($y$) direction.

Calcite dissolution kinetics are more properly expressed in terms of the calcium ion concentration at the fracture surface, $c_s$, rather than the bulk concentration as in Eq.~\eqref{eq:R}:
\begin{equation}
\label{eq:Rsurf}
R(c_s) = \left\{ \begin{array}{cc}
 \!\!\!\!k(c_{sat}-c_s), & c_s<c_{th} \\
 k_n(c_{sat}-c_s)^n, &c_s>c_{th}, \\ 
 \end{array} \right.
\end{equation}
The question then is how to relate the surface concentration, $c_s$, to the bulk concentration, $c$; for this we must consider how diffusion limits the rate of mass transfer from the fracture surface. Diffusive transport of reactant across the fracture aperture can be approximated using an effective mass transfer coefficient or Sherwood number~\citep{Bird2001},
\begin{equation}
\label{eq:Rdiff}
R_{diff} = \frac{D \Sh}{2 h}(c_{s}-c).
\end{equation}
The Sherwood number, $\Sh$, depends on reaction rate but the variation is relatively small~\citep{Hayes1994,Gupta2001}, bounded by two asymptotic limits: constant flux, where $\Sh=8.24$, and constant concentration, where $\Sh=7.54$. In the numerical calculations we used the approximate value $\Sh=8$.

In sufficiently narrow apertures dissolution kinetics are reaction limited, $R \ll R_{diff}$, and $c_s\approx c$ which gives the reactive flux in Eq.~\eqref{eq:R1}. However, as the fracture opens the reaction rate becomes hindered by diffusive transport of reactant across the aperture. Finally, for $k h/D Sh \gg 1$, the dissolution can become entirely diffusion limited, $R_{diff} \ll R$. In that case the surface concentration approaches that of a saturated solution $c_s \approx c_{sat}$ and 
\begin{equation}\label{eq:R2eff}
R_{diff} = \frac{D \Sh}{2 h}(c_{sat}-c).
\end{equation}
Thus the reactive flux is again of the form of Eq.~\eqref{eq:R1} but with an effective rate constant $k_D = D \Sh /2h$. \citet{Weyl1958} assumed diffusion-limited kinetics, but for a typical fracture aperture ($10^{-2} \cm$) $k_D$ is $1-2$ orders of magnitude larger than $k$. In fact reaction-limited kinetics persist, with small modifications from the effects of diffusion, up until fracture apertures of the order of $1\cm$. 

An effective reaction rate, including the effects of diffusion, can be obtained  by balancing the reactive flux, $R$, with the diffusive flux, $R_{diff}$~\citep{Szymczak2009};
\begin{equation}\label{eq:Reff}
\frac{D \Sh}{2h} (c_s - c) = R(c_s).
\end{equation} 
In the case of linear kinetics we then obtain an effective rate constant
\begin{equation}
\label{eq:k_eff}
k_{eff} = \frac{k}{1 + (2kh/D\Sh)},
\end{equation}
which is valid for all apertures. For more reactive materials, such as gypsum, transport corrections are important, even for narrow apertures. In some cases the kinetics is entirely transport limited, as in Eq.~\eqref{eq:R2eff}

In the most general case, non-linear kinetics combined with diffusive mass transfer, we can combine Eqs.~\eqref{eq:Rsurf} and~\eqref{eq:Reff} to obtain the reactive flux in terms of the bulk concentration $c$. Defining the dimensionless quantities:
\begin{equation}
C = \frac{c_{sat} - c}{c_{sat} - c_{th}}, ~~ C_s = \frac{c_{sat} - c_s}{c_{sat} - c_{th}}, ~~ \alpha = \frac{2 k h}{D \Sh},
\end{equation}
Eq.~\eqref{eq:Reff} becomes
\begin{equation}\label{eq:nonlin}
\alpha C_s^n + C_s - C = 0,
\end{equation}
where $C$ is known and $C_s$ is to be found. The numerical simulations use Eq.~\eqref{eq:nonlin}, but the instability analysis (Sec.~\ref{instab}) assumes reaction-limited kinetics, Eq.~\eqref{eq:R}. The difference is small during the initial stages of calcite dissolution.

We used a flux-conserving discretization to solve for the pressure field, which avoids ``numerical saturation''~\citep{Groves1994,Hanna1998}, even when the grid is relatively coarse. The scalar fields, aperture, pressure, and concentration are defined at the nodal points and gradients of the scalar fields are then naturally calculated on the edges of the cells surrounding each node. The divergence of the flux is then automatically calculated at the nodal position. However, the convective flux of reactant was calculated by upwind differencing to ensure stability.

The linear equations for the pressure field were solved using the MUMPS package~\citep{Amestoy2001,Amestoy2006},  and the fluid volume flux was then calculated by differencing. For linear reaction kinetics, the concentration field can be solved directly, but for non-linear kinetics Newton-Raphson iteration was employed, using MUMPS to solve the linear system at each step. Within each iteration of the bulk concentration field, the surface concentrations, Eq.~\eqref{eq:nonlin}, must be determined iteratively as well. The transition between linear ($C_s \ge 1$) and non-linear ($C_s < 1$) kinetics corresponds to $C = 1 + \alpha$, and thus the appropriate branch of Eq.~\eqref{eq:Rsurf}, can be identified from the bulk concentration. Once the concentration field has converged, the aperture field is updated based on the local erosion flux, $2R(c_s)$. We used an explicit midpoint method to determine the time evolution of the aperture field, requiring two complete cycles of the concentration solver at each time step.

\section{Transport and reaction parameters in typical fractures}\label{DaPe}

Fracture apertures are between $0.005\cm$ and $0.1\cm$ \citep{Motyka1984,Paillet1987,Dreybrodt1996}, and hydraulic gradients are of the order of $10^{-3}$ to $10^{-1}$ \citep{Palmer1991,Dijk1998}. This gives a range of characteristic flow velocities in undissolved fractures from $10^{-4} \cm\s^{-1}$ to $1 \cm\s^{-1}$. The corresponding P\'eclet numbers are $0.05 < \Pe < 10^4$, taking the solute diffusion coefficient $D=10^{-5} \mbox{cm}^2\mbox{s}^{-1}$. The dissolution rate for limestone is usually in the range $10^{-5} \cm\s^{-1} - 10^{-4} \cm\s^{-1}$  \citep{Palmer1991,Dreybrodt1996}, which leads to the Damk\"ohler numbers in the range $10^{-5} < \Da < 1$. Thus in most calcite fractures the parameter $\Pe^{-1}\Da$ is small and the diffusive flux can be neglected, as in Sec.~\ref{instab}. Diffusion plays a more prominent role in dissolution of fractured gypsum, where the reaction rates are of the order of $0.01 \cm\s^{-1}$ \citep{Jeschke2001}, and thus $\Pe^{-1}\Da$ is $2-3$ orders of magnitude larger than in limestone systems.

\section{Linear stability analysis}\label{Lsa}

Equation~\eqref{eq:2Dmin} can be non-dimensionalized by scaling length by the penetration length, $x \rightarrow 2k x/q_0$, and time by the time taken to double the initial inlet aperture, $t \rightarrow 2k t c_{sat}/h_0 c_{sol}$. It will be convenient to rewrite the concentration field as a normalized undersaturation, $c \rightarrow (c_{sat} - c)/c_{sat}$. Then, scaling the remaining variables ($h$, $\Vq$, $\nabla p$) by their (constant) initial values ($h_0$, $q_0$, $p_0^\prime$), we can rewrite Eq.~\eqref{eq:2Dmin} in dimensionless form,
\begin{equation}\label{eq:2Dscale}
\begin{array}{cc}
\nabla \cdot \Vq = 0, & \Vq = h^3 \nabla p, \\
\Vq \cdot \nabla c = - c, & \partial_t h = c.
\end{array}
\end{equation}
The one-dimensional solution of~\eqref{eq:2Dscale} is taken as the base state: $h_b(x,t) = 1 + t e^{-x}$, $\Vq_b = \hat \Vx$, and $c_b = e^{-x}$, where ${\hat \Vx}$ indicates a unit vector pointing in the $x$ direction. We consider perturbations about the base state, $c = c_b + \delta c;\ h=h_b + \delta h$ etc. and keep just the linear terms:
\begin{equation}\label{eq:2Dlin}
\begin{array}{cc}
\nabla \cdot \delta \Vq = 0, ~~ \delta \Vq = 3 h_b^{-1} \delta h{\hat \Vx} + h_b^3 \nabla \delta p, \\
\delta q_x \partial_x c_b + \partial_x \delta c = - \delta c, ~~ \partial_t \delta h = \delta c.
\end{array}
\end{equation}

The pressure variation can be eliminated from the equations for the volume flux by cross differentiation,
\begin{equation}
\partial_y \delta q_x - \partial_x \delta q_y = 3 h_b^{-1} \left(\partial_y \delta h - h_b^{\prime} \delta q_y \right).
\end{equation}
where $h_b^\prime = \partial_x h_b$ is the spatial derivative of the base aperture field. Combining this equation with the incompressibility equation, $\partial_y \delta q_y = - \partial_x \delta q_x$ eliminates $\delta q_y$ as well,
\begin{equation}\label{eq:2Dlapl}
(\partial_x^2 + \partial_y^2) \delta q_x = 3 h_b^{-1}\left(\partial_y^2 \delta h + h_b^{\prime} \partial_x \delta q_x \right).
\end{equation}
Finally, rewriting the transport equation from \eqref{eq:2Dlin} in the form $ \delta q_x=\partial_x (e^x \delta c)$ and substituting into \eqref{eq:2Dlapl} leads to equations for the linearized variations in the concentration and aperture fields,
\begin{equation}\label{eq:lin}
\left[h_b (\partial_x^2 + \partial_y^2) - 3 h_b^{\prime} \partial_x \right] \partial_x (e^x \delta c) = 3 \partial_y^2 \delta h, ~~ \partial_t \delta h  = \delta c.
\end{equation}

We now examine the stability of a periodic perturbation in the aperture and concentration fields;
\begin{equation}
\delta h = g(x,t) e^{-x} \sin (u y), ~~ \delta c =\partial_t g(x,t) e^{-x} \sin (u y),
\end{equation}
where the (dimensionless) wavevector $u = 2 \pi l_p/\lambda$ and $\lambda$ is the wavelength of the perturbation. This {\em ansatz} satisfies Eq.~\eqref{eq:lin} provided that
\begin{equation}\label{eq:flin}
\left[h_b  (\partial_x^2 - u^2) - 3 h_b^{\prime}  \partial_x \right] \partial_t \partial_x g + 3  u^2 e^{-x} g = 0.
\end{equation}
In addition to the boundary condition $g(0,t) = 0$, there are boundary conditions from the uniformity of the flow across the inlet and outlet, $\delta q_y (0,y,t) = \delta q_x (x\rightarrow\infty,y,t) = 0$, or
\begin{equation}\label{eq:bc}
\partial_x^2 g(0,t) = 0, ~~ \partial_x g(x\rightarrow\infty,t) = 0.
\end{equation}

Equation~\eqref{eq:flin} does not have an eigensolution because of the time-dependent base state~\citep{Tan1986}, but a numerical solution of Eq.~\eqref{eq:flin} is shown as the solid line in the inset to Fig.~\ref{fig9}. Further analytic progress can be made by freezing the base state at a particular instant of time, $t_0$, replacing $h_b(x,t)$ by $h_b(t_0) = 1 + t_0e^{-x}$ where $t_0$ is then kept constant:
\begin{equation}\label{eq:flin0}
\left[h_b(t_0) (\partial_x^2 - u^2) - 3 h_b^{\prime}(t_0) \partial_x \right] \partial_t \partial_x g + 3 u^2 e^{-x} g = 0.
\end{equation}
In this case an eigensolution can be found of the form
\begin{equation}
\label{eq:feigen}
g(x,t) = {\hat g}(x) e^{\omega t},
\end{equation}
which gives an ordinary differential equation for $\hat g$,
\begin{equation}\label{eq:flint0}
\left[h_b(t_0) (\partial_x^2 - u^2) - 3 h_b^{\prime}(t_0) \partial_x \right] \partial_x {\hat g} + 3 \omega^{-1} u^2 e^{-x} {\hat g} = 0.
\end{equation}
Here we present the solution for the simplest case, $t_0 = 0$, for which $h_b(0)=1$ and $h_b^\prime(0)=0$. A similar but more complex solution can be obtained for arbitrary $t_0$ and results are shown in Fig.~\ref{fig9}.

The general solution of Eq.~\eqref{eq:flint0} with $t_0 = 0$ is
\begin{multline}\label{eq:c}
{\hat g}(x) = A\, _0F_2\left(1+u, 1-u; 3\omega^{-1}u^2e^{-x}\right) \\ + Be^{ux}\, _0F_2\left(1+u, 1-2u; 3\omega^{-1}u^2e^{-x}\right) \\ + Ce^{-ux}\, _0F_2\left(1+u, 1+2u; 3\omega^{-1}u^2e^{-x}\right),
\end{multline}
where $A$, $B$, and $C$ are constants and $_0F_2(p, q; z)$ is a generalized hypergeometric function. The far field boundary condition $\partial_x {\hat g}(x\rightarrow \infty) = 0$ requires that $B = 0$, while the condition ${\hat g}(0) = 0$ is then sufficient to determine the function ${\hat g}(x)$ to within an arbitrary constant, which is the initial amplitude of the perturbation. Imposing the final boundary condition, $\partial_x^2{\hat g}(0) = 0$, gives a dispersion relation for $\omega(u)$,
\begin{multline}\label{eq:disp}
\left[\omega^2 \,
_0\tilde{F}_2\left(1+u,1+2u;3\omega^{-1}u^2\right) +  \right. \\ 3 (1+2u) \omega \,
   _0\tilde{F}_2\left(2+u,2+2u;3\omega^{-1}u^2\right)   \\ \left. +  9 u^2 \, _0\tilde{F}_2\left(3+u,3+2u;3\omega^{-1}u^2\right) \right] {_0\tilde{F}_2}\left(1+u,1-u;3\omega^{-1}u^2\right) = \\ 3
\left[\omega \, _0\tilde{F}_2\left(2+u,2-u;3\omega^{-1}u^2\right) + \right. \\ \left. 3u^2 \, _0\tilde{F}_2\left(3+u,3-u;3\omega^{-1}u^2\right)\right]\,  _0\tilde{F}_2\left(1+u,1+2u;3\omega^{-1}u^2\right), 
\end{multline}
where $_0{\tilde F}_2(p,q;z) = {_0F_2}(p,q;z)/\Gamma(p)\Gamma(q)$. The maximum growth rate (largest positive root) at each $u$ from Eq.~\eqref{eq:disp} is plotted in Fig.~\ref{fig5}.

The derivation of the dispersion relation can be generalized to include lateral diffusion (in the $y$ direction). The convection-diffusion equation, including lateral diffusion and scaled in the same way as Eq.~\eqref{eq:2Dlin}, contains a single parameter, the product $\Pe^{-1}\Da = Dkh_0/q_0^2$,
\begin{equation}\label{eq:2DlinD}
\partial_x (e^x \delta c) - (2\Pe^{-1}\Da) \partial_y^2 (e^x \delta c) = \delta q_x.
\end{equation}
The analogous linear stability analysis leads to the generalization of Eq.~\eqref{eq:flint0} for finite diffusion. Simplifying to the case $t_0=0$,
\begin{equation}\label{eq:flin2}
\left(\partial_x^2 - u^2 \right) \left[ \partial_x + (2\Pe^{-1} \Da) u^2 \right] {\hat g} + 3 \omega^{-1} u^2 e^{-x} {\hat g} = 0.
\end{equation}
which can again be solved in terms of hypergeometric functions. The inset to Fig.~\ref{fig5} shows the dispersion relation for several different values of the product $\Pe^{-1}\Da$. 

At long wavelengths the growth rate is linear in the wavevector, $\omega \rightarrow 3u$, while for small wavelengths an asymptotic analysis, including lateral diffusion, gives $\omega \rightarrow 3\Pe\Da^{-1}u^{-2}$. However, the convective limit (no diffusion) appears to be singular, with $\omega \sim u^{-1}$ at large $u$.


\begin{thebibliography}{42}
\expandafter\ifx\csname natexlab\endcsname\relax\def\natexlab#1{#1}\fi
\expandafter\ifx\csname url\endcsname\relax
  \def\url#1{\texttt{#1}}\fi
\expandafter\ifx\csname urlprefix\endcsname\relax\def\urlprefix{URL }\fi
\providecommand{\eprint}[2][]{\url{#2}}
\providecommand{\bibinfo}[2]{#2}
\ifx\xfnm\relax \def\xfnm[#1]{\unskip,\space#1}\fi
\bibitem[{Amestoy et~al.(2001)Amestoy, Duff, Koster and
  L'Excellent}]{Amestoy2001}
\bibinfo{author}{Amestoy, P.R.}, \bibinfo{author}{Duff, I.S.},
  \bibinfo{author}{Koster, J.}, \bibinfo{author}{L'Excellent, J.Y.},
  \bibinfo{year}{2001}.
\newblock \bibinfo{title}{A fully asynchronous multifrontal solver using
  distributed dynamic scheduling}.
\newblock \bibinfo{journal}{SIAM Journal on Matrix Analysis and Applications}
  \bibinfo{volume}{23}, \bibinfo{pages}{15--41}.
\bibitem[{Amestoy et~al.(2006)Amestoy, Guermouche, L'Excellent and
  Pralet}]{Amestoy2006}
\bibinfo{author}{Amestoy, P.R.}, \bibinfo{author}{Guermouche, A.},
  \bibinfo{author}{L'Excellent, J.Y.}, \bibinfo{author}{Pralet, S.},
  \bibinfo{year}{2006}.
\newblock \bibinfo{title}{Hybrid scheduling for the parallel solution of linear
  systems}.
\newblock \bibinfo{journal}{Parallel Computing} \bibinfo{volume}{32},
  \bibinfo{pages}{136--156}.
\bibitem[{Atkinson(1968)}]{Atkinson1968}
\bibinfo{author}{Atkinson, T.C.}, \bibinfo{year}{1968}.
\newblock \bibinfo{title}{The earliest stage of underground drainage in
  limestone: a speculative discussion}.
\newblock \bibinfo{journal}{Proc. Br. Speleol. Assoc.} \bibinfo{volume}{6},
  \bibinfo{pages}{53–70}.
\bibitem[{Bird et~al.(2001)Bird, Stewart and Lightfoot}]{Bird2001}
\bibinfo{author}{Bird, R.B.}, \bibinfo{author}{Stewart, W.E.},
  \bibinfo{author}{Lightfoot, E.N.}, \bibinfo{year}{2001}.
\newblock \bibinfo{title}{Transport Phenomena}.
\newblock \bibinfo{publisher}{John Wiley \& Sons}, \bibinfo{address}{Department
  of Chemical Engineering, Madison, Wisconsin}.
\bibitem[{Chadam et~al.(1986)Chadam, Hoff, Merino, Ortoleva and
  Sen}]{Chadam1986}
\bibinfo{author}{Chadam, D.}, \bibinfo{author}{Hoff, D.},
  \bibinfo{author}{Merino, E.}, \bibinfo{author}{Ortoleva, P.},
  \bibinfo{author}{Sen, A.}, \bibinfo{year}{1986}.
\newblock \bibinfo{title}{Reactive infiltration instabilities}.
\newblock \bibinfo{journal}{J. Appl. Math.} \bibinfo{volume}{36},
  \bibinfo{pages}{207--221}.
\bibitem[{Cheung and Rajaram(2002)}]{Cheung2002}
\bibinfo{author}{Cheung, W.}, \bibinfo{author}{Rajaram, H.},
  \bibinfo{year}{2002}.
\newblock \bibinfo{title}{Dissolution finger growth in variable aperture
  fractures: {Role} of the tip-region flow field}.
\newblock \bibinfo{journal}{Geophys. Res. Lett.} \bibinfo{volume}{22},
  \bibinfo{pages}{2075}.
\bibitem[{Colombani(2008)}]{Colombani2008}
\bibinfo{author}{Colombani, J.}, \bibinfo{year}{2008}.
\newblock \bibinfo{title}{Measurement of the pure dissolution rate constant of
  a mineral in water}.
\newblock \bibinfo{journal}{Geochim. Cosmochim. Acta} \bibinfo{volume}{72},
  \bibinfo{pages}{15634–15640}.
\bibitem[{Colombani and Bert(2007)}]{Colombani2007}
\bibinfo{author}{Colombani, J.}, \bibinfo{author}{Bert, J.},
  \bibinfo{year}{2007}.
\newblock \bibinfo{title}{Holographic interferometry study of the dissolution
  and diffusion of gypsum in water}.
\newblock \bibinfo{journal}{Geochim. Cosmochim. Acta} \bibinfo{volume}{71},
  \bibinfo{pages}{1913--1920}.
\bibitem[{Detwiler et~al.(2003)Detwiler, Glass and Bourcier}]{Detwiler2003}
\bibinfo{author}{Detwiler, R.L.}, \bibinfo{author}{Glass, R.J.},
  \bibinfo{author}{Bourcier, W.L.}, \bibinfo{year}{2003}.
\newblock \bibinfo{title}{Experimental observations of fracture dissolution:
  {The} role of {P\'eclet} number in evolving aperture variability}.
\newblock \bibinfo{journal}{Geophys. Res. Lett.} \bibinfo{volume}{30},
  \bibinfo{pages}{1648}.
\bibitem[{Detwiler and Rajaram(2007)}]{Detwiler2007}
\bibinfo{author}{Detwiler, R.L.}, \bibinfo{author}{Rajaram, H.},
  \bibinfo{year}{2007}.
\newblock \bibinfo{title}{Predicting dissolution patterns in variable aperture
  fractures: Evaluation of an enhanced depth-averaged computational model}.
\newblock \bibinfo{journal}{Water Resourc. Res.} \bibinfo{volume}{43},
  \bibinfo{pages}{W04403}.
\bibitem[{Dijk and Berkowitz(1998)}]{Dijk1998}
\bibinfo{author}{Dijk, P.}, \bibinfo{author}{Berkowitz, B.},
  \bibinfo{year}{1998}.
\newblock \bibinfo{title}{Precipitation and dissolution of reactive solutes in
  fractures}.
\newblock \bibinfo{journal}{Water Resourc. Res.} \bibinfo{volume}{34},
  \bibinfo{pages}{457--470}.
\bibitem[{Dreybrodt(1990)}]{Dreybrodt1990}
\bibinfo{author}{Dreybrodt, W.}, \bibinfo{year}{1990}.
\newblock \bibinfo{title}{The role of dissolution kinetics in the development
  of karst aquifers in limestone: \uppercase{A} model simulation of karst
  evolution,}.
\newblock \bibinfo{journal}{Water Resourc. Res.} \bibinfo{volume}{98},
  \bibinfo{pages}{639--655}.
\bibitem[{Dreybrodt(1996)}]{Dreybrodt1996}
\bibinfo{author}{Dreybrodt, W.}, \bibinfo{year}{1996}.
\newblock \bibinfo{title}{Principles of early development of karst conduits
  under natural and man-made conditions revealed by mathematical analysis of
  numerical models}.
\newblock \bibinfo{journal}{Water Resourc. Res.} \bibinfo{volume}{32},
  \bibinfo{pages}{2923--2935}.
\bibitem[{Dreybrodt et~al.(2002)Dreybrodt, Romanov and
  Gabrov\u{s}ek}]{Dreybrodt2002}
\bibinfo{author}{Dreybrodt, W.}, \bibinfo{author}{Romanov, D.},
  \bibinfo{author}{Gabrov\u{s}ek, F.}, \bibinfo{year}{2002}.
\newblock \bibinfo{title}{Karstification below dam sites: a model of increasing
  leakage from reservoirs}.
\newblock \bibinfo{journal}{Environ. Geol.} \bibinfo{volume}{42},
  \bibinfo{pages}{518--524}.
\bibitem[{Durham et~al.(2001)Durham, Bourcier and Burton}]{Durham2001}
\bibinfo{author}{Durham, W.B.}, \bibinfo{author}{Bourcier, W.L.},
  \bibinfo{author}{Burton, E.A.}, \bibinfo{year}{2001}.
\newblock \bibinfo{title}{Direct observation of reactive flow in a single
  fracture}.
\newblock \bibinfo{journal}{Water Resourc. Res.} \bibinfo{volume}{37},
  \bibinfo{pages}{1--12}.
\bibitem[{Farrell and Ioannou(1996)}]{Farrell1996}
\bibinfo{author}{Farrell, B.F.}, \bibinfo{author}{Ioannou, P.J.},
  \bibinfo{year}{1996}.
\newblock \bibinfo{title}{Generalized stability theory. {Part II}:
  {Nonautonomous} operators}.
\newblock \bibinfo{journal}{J. Atmos. Sci.} \bibinfo{volume}{53},
  \bibinfo{pages}{2041--2053}.
\bibitem[{Fredd and Fogler(1998)}]{Fredd1998}
\bibinfo{author}{Fredd, C.N.}, \bibinfo{author}{Fogler, H.S.},
  \bibinfo{year}{1998}.
\newblock \bibinfo{title}{Influence of transport and reaction on wormhole
  formation in porous media}.
\newblock \bibinfo{journal}{AIChE J.} \bibinfo{volume}{44},
  \bibinfo{pages}{1933--1949}.
\bibitem[{Gouze et~al.(2003)Gouze, Noiriel, Bruderer and Loggia}]{Gouze2003}
\bibinfo{author}{Gouze, P.}, \bibinfo{author}{Noiriel, C.},
  \bibinfo{author}{Bruderer, C.}, \bibinfo{author}{Loggia, D.},
  \bibinfo{year}{2003}.
\newblock \bibinfo{title}{X-ray tomography characterization of fracture
  surfaces during dissolution}.
\newblock \bibinfo{journal}{Geophys. Res. Lett.} \bibinfo{volume}{30},
  \bibinfo{pages}{1267}.
\bibitem[{Groves and Howard(1994)}]{Groves1994}
\bibinfo{author}{Groves, C.G.}, \bibinfo{author}{Howard, A.D.},
  \bibinfo{year}{1994}.
\newblock \bibinfo{title}{Minimum hydrochemical conditions allowing limestone
  cave development}.
\newblock \bibinfo{journal}{Water Resourc. Res.} \bibinfo{volume}{30},
  \bibinfo{pages}{607--615}.
\bibitem[{Gubiec and Szymczak(2008)}]{Gubiec2008}
\bibinfo{author}{Gubiec, T.}, \bibinfo{author}{Szymczak, P.},
  \bibinfo{year}{2008}.
\newblock \bibinfo{title}{Fingered growth in channel geometry: A
  \uppercase{L}oewner-equation approach}.
\newblock \bibinfo{journal}{Phys. Rev. E} \bibinfo{volume}{77},
  \bibinfo{pages}{041602}.
\bibitem[{Gupta and Balakotaiah(2001)}]{Gupta2001}
\bibinfo{author}{Gupta, N.}, \bibinfo{author}{Balakotaiah, V.},
  \bibinfo{year}{2001}.
\newblock \bibinfo{title}{Heat and mass transfer coefficients in catalytic
  monoliths}.
\newblock \bibinfo{journal}{Chem. Eng. Sci.} \bibinfo{volume}{56},
  \bibinfo{pages}{4771--4786}.
\bibitem[{Hanna and Rajaram(1998)}]{Hanna1998}
\bibinfo{author}{Hanna, R.B.}, \bibinfo{author}{Rajaram, H.},
  \bibinfo{year}{1998}.
\newblock \bibinfo{title}{Influence of aperture variability on dissolutional
  growth of fissures in karst formations}.
\newblock \bibinfo{journal}{Water Resourc. Res.} \bibinfo{volume}{34},
  \bibinfo{pages}{2843--2853}.
\bibitem[{Hayes and Kolaczkowski(1994)}]{Hayes1994}
\bibinfo{author}{Hayes, R.E.}, \bibinfo{author}{Kolaczkowski, S.T.},
  \bibinfo{year}{1994}.
\newblock \bibinfo{title}{Mass and heat transfer effects in catalytic monolith
  reactors}.
\newblock \bibinfo{journal}{Chem. Eng. Sci.} \bibinfo{volume}{49},
  \bibinfo{pages}{3587--3599}.
\bibitem[{Hinch and Bhatt(1990)}]{Hinch1990}
\bibinfo{author}{Hinch, E.J.}, \bibinfo{author}{Bhatt, B.S.},
  \bibinfo{year}{1990}.
\newblock \bibinfo{title}{Stability of an acid front moving through porous
  rock}.
\newblock \bibinfo{journal}{J. Fluid Mech.} \bibinfo{volume}{212},
  \bibinfo{pages}{279--288}.
\bibitem[{Jeschke et~al.(2001)Jeschke, Vosbeck and Dreybrodt}]{Jeschke2001}
\bibinfo{author}{Jeschke, A.A.}, \bibinfo{author}{Vosbeck, K.},
  \bibinfo{author}{Dreybrodt, W.}, \bibinfo{year}{2001}.
\newblock \bibinfo{title}{Surface controlled dissolution rates of gypsum in
  aqueous solutions exhibit nonlinear dissolution kinetics}.
\newblock \bibinfo{journal}{Geochim. Cosmochim. Acta}
  \bibinfo{volume}{65}, \bibinfo{pages}{27--34}.
\bibitem[{Long et~al.(1982)Long, Remer, Wilson and Witherspoon}]{Long1982}
\bibinfo{author}{Long, J.C.S.}, \bibinfo{author}{Remer, J.S.},
  \bibinfo{author}{Wilson, C.R.}, \bibinfo{author}{Witherspoon, P.A.},
  \bibinfo{year}{1982}.
\newblock \bibinfo{title}{Porous-media equivalents for networks of
  discontinuous fractures}.
\newblock \bibinfo{journal}{Water Resour. Res.} \bibinfo{volume}{18},
  \bibinfo{pages}{645--658}.
\bibitem[{Motyka and Wilk(1984)}]{Motyka1984}
\bibinfo{author}{Motyka, I.}, \bibinfo{author}{Wilk, Z.}, \bibinfo{year}{1984}.
\newblock \bibinfo{title}{Hydraulic structure of karst-fissured {Triassic}
  rocks in the vicinity of {Olkusz} {(Poland)}}.
\newblock \bibinfo{journal}{Kras i Speleologia} \bibinfo{volume}{14},
  \bibinfo{pages}{11--24}.
\bibitem[{Paillet et~al.(1987)Paillet, Hess, Cheng and Harding}]{Paillet1987}
\bibinfo{author}{Paillet, F.L.}, \bibinfo{author}{Hess, A.E.},
  \bibinfo{author}{Cheng, C.H.}, \bibinfo{author}{Harding, E.},
  \bibinfo{year}{1987}.
\newblock \bibinfo{title}{Characterization of fracture permeability with
  high-resolution vertical flow measurements during borehole pumping}.
\newblock \bibinfo{journal}{Ground Water} \bibinfo{volume}{25},
  \bibinfo{pages}{28–40}.
\bibitem[{Palmer(1991)}]{Palmer1991}
\bibinfo{author}{Palmer, A.N.}, \bibinfo{year}{1991}.
\newblock \bibinfo{title}{Origin and morphology of limestone caves}.
\newblock \bibinfo{journal}{GSA Bulletin} \bibinfo{volume}{103},
  \bibinfo{pages}{1--21}.
\bibitem[{Plummer and Wigley(1976)}]{Plummer1976}
\bibinfo{author}{Plummer, L.N.}, \bibinfo{author}{Wigley, T.L.M.},
  \bibinfo{year}{1976}.
\newblock \bibinfo{title}{The dissolution of calcite in {CO2}-water systems at
  $25\,^{\circ}\mathrm{C}$ and 1 atmosphere total pressure}.
\newblock \bibinfo{journal}{Geochim. Cosmoch. Acta} \bibinfo{volume}{40},
  \bibinfo{pages}{191--202}.
\bibitem[{Raines and Dewers(1997)}]{Raines1997}
\bibinfo{author}{Raines, M.A.}, \bibinfo{author}{Dewers, T.A.},
  \bibinfo{year}{1997}.
\newblock \bibinfo{title}{Mixed transport/reaction control of gypsum
  dissolution lunetics in aqueous solutions and initiation of gypsum karst}.
\newblock \bibinfo{journal}{Chem. Geol.} \bibinfo{volume}{140},
  \bibinfo{pages}{29--48}.
\bibitem[{Romanov et~al.(2003)Romanov, Gabrov\u{s}ek and
  Dreybrodt}]{Romanov2003}
\bibinfo{author}{Romanov, D.}, \bibinfo{author}{Gabrov\u{s}ek, F.},
  \bibinfo{author}{Dreybrodt, W.}, \bibinfo{year}{2003}.
\newblock \bibinfo{title}{Dam sites in soluble rocks: a model of increasing
  leakage by dissolutional widening of fractures beneath a dam}.
\newblock \bibinfo{journal}{Eng. Geol.} \bibinfo{volume}{70},
  \bibinfo{pages}{129--145}.
\bibitem[{Shaw(2000)}]{Shaw2000}
\bibinfo{author}{Shaw, T.R.}, \bibinfo{year}{2000}.
\newblock \bibinfo{title}{Speleogenesis: Evolution of karst aquifers}.
  \bibinfo{publisher}{Natl. Speleol. Soc., Huntsville}.
\newblock pp. \bibinfo{pages}{60--73}.
\bibitem[{Sherwood(1987)}]{Sherwood1987}
\bibinfo{author}{Sherwood, J.D.}, \bibinfo{year}{1987}.
\newblock \bibinfo{title}{Stability of a plane reaction front in a porous
  medium}.
\newblock \bibinfo{journal}{Chem. Eng. Sci.} \bibinfo{volume}{42},
  \bibinfo{pages}{1823--1829}.
\bibitem[{Siemers and Dreybrodt(1998)}]{Siemers1998}
\bibinfo{author}{Siemers, J.}, \bibinfo{author}{Dreybrodt, W.},
  \bibinfo{year}{1998}.
\newblock \bibinfo{title}{Early development of karst aquifers on percolation
  networks of fractures in limestone}.
\newblock \bibinfo{journal}{Water Resourc. Res.} \bibinfo{volume}{34},
  \bibinfo{pages}{409--419}.
\bibitem[{Svensson and Dreybrodt(1992)}]{Svensson1992}
\bibinfo{author}{Svensson, U.}, \bibinfo{author}{Dreybrodt, W.},
  \bibinfo{year}{1992}.
\newblock \bibinfo{title}{Dissolution kinetics of natural calcite minerals in
  {CO2}-water systems approaching calcite equilibrium}.
\newblock \bibinfo{journal}{Chem. Geol.} \bibinfo{volume}{100},
  \bibinfo{pages}{129--145}.
\bibitem[{Szymczak and Ladd(2006)}]{Szymczak2006}
\bibinfo{author}{Szymczak, P.}, \bibinfo{author}{Ladd, A.J.C.},
  \bibinfo{year}{2006}.
\newblock \bibinfo{title}{A network model of channel competition in fracture
  dissolution}.
\newblock \bibinfo{journal}{Geophys. Res. Lett.} \bibinfo{volume}{33},
  \bibinfo{pages}{L05401}.
\bibitem[{Szymczak and Ladd(2009)}]{Szymczak2009}
\bibinfo{author}{Szymczak, P.}, \bibinfo{author}{Ladd, A.J.C.},
  \bibinfo{year}{2009}.
\newblock \bibinfo{title}{Wormhole formation in dissolving fractures}.
\newblock \bibinfo{journal}{J. Geophys. Res.} \bibinfo{volume}{114},
  \bibinfo{pages}{B06203}.
\bibitem[{Tan and Homsy(1986)}]{Tan1986}
\bibinfo{author}{Tan, C.T.}, \bibinfo{author}{Homsy, G.M.},
  \bibinfo{year}{1986}.
\newblock \bibinfo{title}{Stability of miscible displacement in porous media:
  {Rectilinear} flow.}
\newblock \bibinfo{journal}{Phys. Fluids} \bibinfo{volume}{29},
  \bibinfo{pages}{3549--3556}.
\bibitem[{Weyl(1958)}]{Weyl1958}
\bibinfo{author}{Weyl, P.K.}, \bibinfo{year}{1958}.
\newblock \bibinfo{title}{The solution kinetics of calcite}.
\newblock \bibinfo{journal}{J. Geol.} \bibinfo{volume}{66},
  \bibinfo{pages}{163--176}.
\bibitem[{White(1977)}]{White1977}
\bibinfo{author}{White, W.B.}, \bibinfo{year}{1977}.
\newblock \bibinfo{title}{Role of solution kinetics in the development of karst
  aquifers}.
\newblock \bibinfo{journal}{Mem Int Assoc Hydrogeol} \bibinfo{volume}{12},
  \bibinfo{pages}{503--517}.
\bibitem[{White and Longyear(1962)}]{White1962}
\bibinfo{author}{White, W.B.}, \bibinfo{author}{Longyear, J.},
  \bibinfo{year}{1962}.
\newblock \bibinfo{title}{Some limitations on speleogenetic speculation imposed
  by the hydraulics of groundwater flow in limestone}.
\newblock \bibinfo{journal}{Nittany Grotto Newl.} \bibinfo{volume}{10},
  \bibinfo{pages}{155--167}.

\end{thebibliography}
\end{document}